\documentclass[aps,pra,twocolumn,showpacs]{revtex4}

\usepackage{graphicx}
\newcommand{\tr}{\mbox{Tr} }
\newcommand{\ket}[1]{\left | #1 \right \rangle}
\newcommand{\bra}[1]{\left \langle #1 \right |}

\newcommand{\hilbert}{{\cal H}}
\begin{document}
\title{Local vertical measurements and violation of Bell inequality}
\author{Yang Xiang}
\email{njuxy@sina.com}
\author{Shi-Jie Xiong }

\affiliation{National Laboratory of Solid State Microstructures and
Department of Physics, Nanjing University, Nanjing 210093, China}
\date{\today}
\begin{abstract}

For two qubits belonging to Alice and Bob, we derive an approach to
setup the bound of Bell operator in the condition that Alice and Bob
continue to perform local vertical measurements. For pure states we
find that if the entanglement of the two qubits is less than
$0.2644$ (measured with von Neumann entropy) the violation of the
Bell inequality will never be realized, and only when the
entanglement is equal to $1$ the maximal violation ($2\sqrt{2}$) can
occur. For specific form of mixed states, we prove that the bound of
the Bell inequality depends on the concurrence. Only when the
concurrence is greater than $0.6$ the violation of the Bell
inequality can occur, and the maximal violation can never be
achieved. We suggest that the bound of the Bell operator in the
condition of local vertical measurements may be used as a measure of
the entanglement.

\end{abstract}

 \maketitle




The local realism theory (LRT) \cite{einstein} states that physical
systems can be described by local objective properties (physical
reality) which are independent of observation. The Bell inequality
\cite{bell}, however, sets bound for correlations of local
observables within any LRT. The violation of the Bell inequality
means that the quantum mechanics cannot be regarded as a LRT. There
is profound relation between the violation of the Bell inequality
and the quantum entanglement, and this relation has been formulated
as the entanglement witness \cite{ew}. As early as 1991, Gisin
\textit{et al.} \cite{gisin1} pointed out that the Bell inequality
is satisfied for any separable quantum state, but may be violated by
any purely entangled state if one chooses a proper measurement
setting.

The original Bell inequality has been extended to a more general
inequality by Clauser, Horne, Shimony, and Holt (CHSH inequality)
\cite{chsh}. Consider a bipartite quantum system including qubit $a$
belonging to Alice and and qubit $b$ belonging to Bob. Alice and Bob
are at distant sites and choose to measure one of two dichotomous
observables: $A$ or $A'$ at qubit $a$ and $B$ or $B'$ at qubit $b$.
All observables have the spectrum in $\{-1,1\}$. In this Letter we
only consider traceless spin observables, e.g.,
$A=\bf{a}\cdot\bf{\sigma}$ and analogously for $A', B, B'$. There is
a so-called Bell operator \cite{belloperator},
\begin{eqnarray}
W\equiv A\otimes(B+B')+A'\otimes(B-B'). \label{belloperator}
\end{eqnarray}
The CHSH inequality is
\begin{eqnarray}
|\langle W\rangle_{\rho}|\leq 2, \label{chsh}
\end{eqnarray}
where $\langle W\rangle_{\rho}$ is the expected value of $W$ in
state $\rho$. For any quantum states, a bound of $W$ is given by the
Tsirelson inequality \cite{tsir,landau}
\begin{eqnarray}
|\langle W\rangle_{\rho}|\leq
\sqrt{4+|\langle[A,A']\otimes[B,B']\rangle_{\rho}|}. \label{tsir}
\end{eqnarray}
Landau \cite{landau} has  pointed out that the Tsirelson inequality
is tight, i.e., for any choices of the observables, there exists a
state $\rho$ which can make
\begin{eqnarray}
\max_{\rho\in D}|\langle W\rangle_{\rho}|=
\sqrt{4+|\langle[A,A']\otimes[B,B']\rangle_{\rho}|}, \label{landau}
\end{eqnarray}
where $D$ is the set of all quantum states. Tsirelson \cite{tsir}
has proved that for spin observables $\max_{\rho\in D}|\langle
W\rangle_{\rho}|$ can be obtained in a pure two-qubit state. From
the Tsirelson inequality it is clear that if one wants to produce a
violation of the CHSH inequality he must carry out measurements on
pairs of non-commuting spin observables for both particles, and if
one wants to achieve the maximal violation ($2\sqrt{2}$) allowed by
the quantum theory he has to choose both pairs of local observables
to be anti-commuting. The latter corresponds to the case that both
Alice and Bob carry out vertical measurements,
$\bf{a}\cdot\bf{a'}=\bf{b}\cdot\bf{b'}=0$. In a recent work
\cite{seev}, Seevinck and Uffink have proved that for entangled
state if both the local angles,
$\theta_{a}=\arccos(\bf{a}\cdot\bf{a'})$ and
$\theta_{b}=\arccos(\bf{b}\cdot\bf{b'})$, increase from zero to
$\pi/2$ the maximal violation of the CHSH inequality increases.

In this Letter we will investigate the following question: What is
the bound of $|\langle W\rangle_{\rho}|$ for a given state $\rho$ in
the condition of local vertical measurements? For this purpose we
derive the analytical expression of the tight upper bound of
$|\langle W\rangle_{\rho}|$ for any given pure state $\rho$, and
show that if Alice and Bob both perform vertical measurements they
would never find violation of CHSH inequality if $\rho$ is not an
``enough'' entangled state. We also derive an approach which can be
used to deal with the case of mixed states. For mixed states of a
specific form we calculate the bound and find that it depends on the
concurrence. We argue that this bound of the Bell operator in the
condition of local vertical measurements can be used as a measure of
the quantum entanglement for any states.

We assume that
\begin{eqnarray}
&&A=\sigma_{z}^{a}~~~~~~~~~~~~~~~~~A'=\sigma_{x}^{a}\nonumber\\
&&B=\frac{-\sigma_{z}^{b}-\sigma_{x}^{b}}{\sqrt{2}}~~~~~~~B'=\frac{\sigma_{z}^{b}-\sigma_{x}^{b}}{\sqrt{2}},
\label{ab}
\end{eqnarray}
thus $W$ can be written as
\begin{eqnarray}
W&\equiv& A\otimes(B+B')+A'\otimes(B-B')\nonumber\\
&=&\left(\begin{array}{c}
-\sqrt{2}~~~~~~~0~~~~~~~0~~~~~~~-\sqrt{2}\\
~0~~~~~~~\sqrt{2}~~~-\sqrt{2}~~~~~~~~~0\\
~0~~~~~-\sqrt{2}~~~~~\sqrt{2}~~~~~~~~~0\\
-\sqrt{2}~~~~~~~0~~~~~~~~0~~~~~-\sqrt{2}
\end{array}\right).\label{w}
\end{eqnarray}
So an arbitrary local vertical measurement scheme can be written as
$(U^{a}\otimes U^{b})^{\dag}W(U^{a}\otimes U^{b})$, where $U^{a(b)}$
is an arbitrary unitary operation on $a(b)$. For a given state
$\rho$, we have to find some $U^{a}$ and $U^{b}$ such that $|\langle
(U^{a}\otimes U^{b})^{\dag}W(U^{a}\otimes U^{b})\rangle_{\rho}|$
takes its maximum value.

\begin{figure}[h]
\includegraphics[width=0.8\columnwidth,
height=0.6\columnwidth]{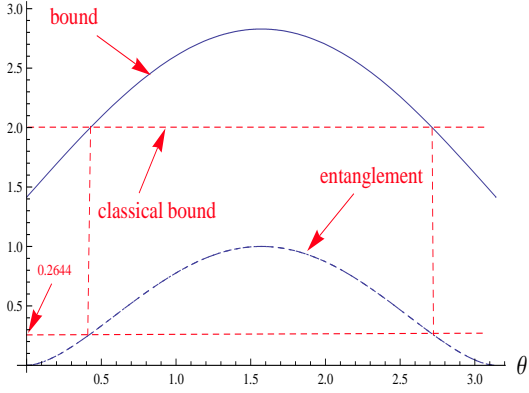} \caption{The bound of $|\langle
U^{\dag}W U\rangle_{\psi}|$ (solid line) for any $\psi$. When this
bound is greater than the classical bound (dashed line), Alice and
Bob achieve the violation of the CHSH inequality. The entanglement
(dot-dashed line) of $\psi$ is also shown.} \label{fig1}
\end{figure}

We first discuss the case of $\rho$ being a pure state. The
eigenvalues and eigenvectors of $W$ are
\begin{eqnarray}
&&-2\sqrt{2}\longleftrightarrow\eta_{1}=
\frac{1}{\sqrt{2}}\left(\begin{array}{c}1\\0\\0\\1\end{array}\right),
0\longleftrightarrow\eta_{2}=\frac{1}{\sqrt{2}} \left(\begin{array}{c}-1\\0\\0\\1\end{array}\right)
\nonumber\\
&&2\sqrt{2}\longleftrightarrow\eta_{3}=\frac{1}{\sqrt{2}}
\left(\begin{array}{c}0\\-1\\1\\0\end{array}\right),
~0\longleftrightarrow\eta_{4}=\frac{1}{\sqrt{2}}\left(\begin{array}{c}0\\
1\\1\\0\end{array}\right).\nonumber\\
\label{eigen}
\end{eqnarray}
We find that the Hilbert space of the two qubits, $\hilbert
=C^{2}\otimes C^{2}$, can be divided into two disjoint subspaces:
$\hilbert_{1}$($\{\eta_{1},\eta_{2}\}$) and
$\hilbert_{2}$($\{\eta_{3},\eta_{4}\}$). It is obvious that in order
to get the maximum of $|\langle (U^{a}\otimes
U^{b})^{\dag}W(U^{a}\otimes U^{b})\rangle_{\rho}|$ one must find
some $U^{a}\otimes U^{b}$ which can map $\psi$ into either
$\hilbert_{1}$ or $\hilbert_{2}$. Here $\rho=\ket{\psi}\bra{\psi}$.
We consider a general pure state
\begin{eqnarray}
\ket{\psi}&=&\cos(\frac{\theta}{2})\ket{{\bf n}}_{a}\ket{{\bf
m}}_{b}\nonumber\\
&~~~~&+e^{i\chi}\sin(\frac{\theta}{2})\ket{{\bf -n}}_{a}\ket{{\bf
-m}}_{b}, \label{www}
\end{eqnarray}
where ${\bf n}$ and ${\bf m}$ are two points on the Poincar\'{e}
sphere, and the subscript specifies the related qubit $a$ or $b$.
The ``angle'' $\theta$ in Eq. (\ref{www}) determines the degree of
entanglement in the state. The angle satisfies $0\leq\theta\leq\pi$,
$\theta=0$ and $\theta=\pi$ correspond to the product states, and
the maximal entanglement occurs at $\theta=\frac{\pi}{2}$. We can
apply a proper product unitary operation on $\psi$ such that
\begin{eqnarray}
\ket{\psi}&=&\cos\left(\frac{\theta}{2}\right)\ket{+z}_{A}\ket{-z}_{B}\nonumber\\
&&+e^{i\chi}\sin\left(\frac{\theta}{2}\right)\ket{-z}_{A}\ket{+z}_{B}\nonumber\\
&=&(0,~~~\cos(\theta/2),~~~e^{i\chi}\sin(\theta/2),~~~0)^{T}.
\label{state}
\end{eqnarray}
An arbitrary unitary operation on a single qubit can be written as
\cite{nielsen}
\begin{eqnarray}
U&\equiv&U(\alpha,\beta,\gamma,\delta)\nonumber\\
&=&e^{-i \alpha}R_{z}(\beta)R_{y}(\gamma)R_{z}(\delta)\nonumber\\
\nonumber\\
&=&e^{-i \alpha}\left(
\begin{array}{c}
e^{i (-\beta/2-\delta/2)}\cos\frac{\gamma}{2}~~~~-e^{i (-\beta/2+\delta/2)}\sin\frac{\gamma}{2}\\
\\
e^{i (+\beta/2-\delta/2)}\sin\frac{\gamma}{2}~~~~~~e^{i
(+\beta/2+\delta/2)}\cos\frac{\gamma}{2}
\end{array}\right), \nonumber\\
\label{uuu}
\end{eqnarray}
where $\alpha, \beta, \gamma$ and $\delta$ are real numbers, and
$R_{y(z)}$ is the rotation operator about the $y(z)$ axis. We use
$U^{a}(\alpha,\beta,\gamma,\delta)$($U^{b}(\alpha',\beta',\gamma',\delta')$)
to denote an arbitrary unitary operation on $a$($b$). We can express
$U^{a}\otimes U^{b}$ as follows:
\begin{widetext}
\begin{eqnarray}
U^{a}\otimes U^{b}=\left(\begin{array}{c}
e^{i\xi_{11}}\cos(\gamma/2)\cos(\gamma'/2)~~-e^{i\xi_{12}}\cos(\gamma/2)\sin(\gamma'/2)~~
-e^{i\xi_{13}}\sin(\gamma/2)\cos(\gamma'/2)~~e^{i\xi_{14}}\sin(\gamma/2)\sin(\gamma'/2)\\
e^{i\xi_{21}}\cos(\gamma/2)\sin(\gamma'/2)~~~~e^{i\xi_{22}}\cos(\gamma/2)\cos(\gamma'/2)
~~~-e^{i\xi_{23}}\sin(\gamma/2)\sin(\gamma'/2)~-e^{i\xi_{24}}\sin(\gamma/2)\cos(\gamma'/2)\\
e^{i\xi_{31}}\sin(\gamma/2)\cos(\gamma'/2)~~-e^{i\xi_{32}}\sin(\gamma/2)\sin(\gamma'/2)~~
e^{i\xi_{33}}\cos(\gamma/2)\cos(\gamma'/2)~~-e^{i\xi_{34}}\cos(\gamma/2)\sin(\gamma'/2)\\
e^{i\xi_{41}}\sin(\gamma/2)\sin(\gamma'/2)~~~~~e^{i\xi_{42}}\sin(\gamma/2)\cos(\gamma'/2)~~~~
e^{i\xi_{43}}\cos(\gamma/2)\sin(\gamma'/2)~~~~~e^{i\xi_{44}}\cos(\gamma/2)\cos(\gamma'/2)
\end{array}\right), \nonumber
\end{eqnarray}
\end{widetext}
where all $\xi_{ij}$ are related to tunable parameters $\alpha,
\alpha', \beta, \beta', \delta, \delta'$ from Eq. (\ref{uuu}). Now
we map $\ket{\psi}$ into subspace $\hilbert_{1}$ by choosing
suitable $U^{a}\otimes U^{b}$. We denote $\ket{\psi'}=U^{a}\otimes
U^{b}\ket{\psi}$. It is found that we have to choose $\gamma$ and
$\gamma'$ in such a way that $\cos(\gamma'/2)=\sin(\gamma/2)=0$ or
$\sin(\gamma'/2)=\cos(\gamma/2)=0$, because otherwise $\psi'$ will
have component state which is in $\hilbert_{2}$ and this will reduce
$|\bra{\psi'}W\ket{\psi'}|$. When
$\cos(\gamma'/2)=\sin(\gamma/2)=0$, we can obtain
\begin{eqnarray}
\psi'&=&U^{a}\otimes U^{b}\ket{\psi}\nonumber\\
&=&\left(-e^{i\xi_{12}}\cos(\theta/2)~~~~~0~~~~~0~~~~~
e^{i(\xi_{43}+\chi)}\sin(\theta/2)\right)^{T}.\nonumber\\
\label{state2}
\end{eqnarray}
Then $|\langle (U^{a}\otimes U^{b})^{\dag}W(U^{a}\otimes
U^{b})\rangle_{\rho}|$ can be obtained as follows
\begin{eqnarray}
&&|\langle (U^{a}\otimes U^{b})^{\dag}W(U^{a}\otimes
U^{b})\rangle_{\psi}|\nonumber\\
&=&|\bra{\psi'}W\ket{\psi'}|\nonumber\\
&=&|-2\sqrt{2}|\cdot |\langle\eta_{1}|\psi'\rangle|^{2}\nonumber\\
&=&\sqrt{2}\cdot
\big|-e^{i\xi_{12}}\cos(\theta/2)+e^{i(\xi_{43}+\chi)}\sin(\theta/2)\big|^2.
\end{eqnarray}
Since $0\leq\theta\leq\pi$, we can take
$-e^{i\xi_{12}}=e^{i(\xi_{43}+\chi)}=0$, and get the maximum value
of $|\langle (U^{a}\otimes U^{b})^{\dag}W(U^{a}\otimes
U^{b})\rangle_{\rho}|$ as
\begin{eqnarray}
|\langle (U^{a}\otimes U^{b})^{\dag}W(U^{a}\otimes
U^{b})\rangle_{\psi}|_{max}=\sqrt{2}\cdot(\sin\theta+1).
\label{outcome1}
\end{eqnarray}

When $\sin(\gamma'/2)=\cos(\gamma/2)=0$, we have
\begin{eqnarray}
\psi'&=&U^{a}\otimes U^{b}\ket{\psi}\nonumber\\
&=&\left(-e^{i(\xi_{13}+\chi)}\sin(\theta/2)~~~~~0~~~~~0~~~~~e^{i\xi_{42}}\cos(\theta/2)
\right)^{T}, \nonumber\\
\label{state3}
\end{eqnarray}
and $|\langle (U^{a}\otimes U^{b})^{\dag}W(U^{a}\otimes
U^{b})\rangle_{\rho}|$ can be calculated as
\begin{eqnarray}
&&|\langle (U^{a}\otimes U^{b})^{\dag}W(U^{a}\otimes
U^{b})\rangle_{\psi}|\nonumber\\
&=&|\bra{\psi'}W\ket{\psi'}|\nonumber\\
&=&|-2\sqrt{2}|\cdot |\langle\eta_{1}|\psi'\rangle|^{2}\nonumber\\
&=&\sqrt{2}\cdot
\big|-e^{i(\xi_{13}+\chi)}\sin(\theta/2)+e^{i\xi_{42}}\cos(\theta/2)\big|^2.
\end{eqnarray}
Since $0\leq\theta\leq\pi$, we take
$-e^{i(\xi_{13}+\chi)}=e^{i\xi_{42}}=0$ and obtain the same maximum
value of $|\langle (U^{a}\otimes U^{b})^{\dag}W(U^{a}\otimes
U^{b})\rangle_{\rho}|$.

We can also map $\psi$ into subspace $\hilbert_{2}$ and obtain the
same maximum of $|\langle (U^{a}\otimes U^{b})^{\dag}W(U^{a}\otimes
U^{b})\rangle_{\psi}|$ in a similar way.

In Fig. \ref{fig1} we plot the maximum of $|\langle (U^{a}\otimes
U^{b})^{\dag}W(U^{a}\otimes U^{b})\rangle_{\psi}|$, which we call as
``bound'', and the entanglement of $\psi$, which is calculated by
using the von Neumann entropy, as functions of $\theta$. We find
that when $\sin(\theta)\leq\sqrt{2}-1$ and Alice and Bob continue to
perform local vertical measurements, they will never reach the
violation of the CHSH inequality. According to the Tsirelson
inequality, if one wants to achieve the maximal violation allowed by
the quantum theory he has to properly choose both pairs of local
vertical measurements. So from Fig. \ref{fig1} we can see that one
cannot get the maximal violation ($2\sqrt{2}$) of the CHSH
inequality unless $\psi$ is a maximally entangled state.

\begin{figure}[h]
\includegraphics[width=0.8\columnwidth,
height=0.6\columnwidth]{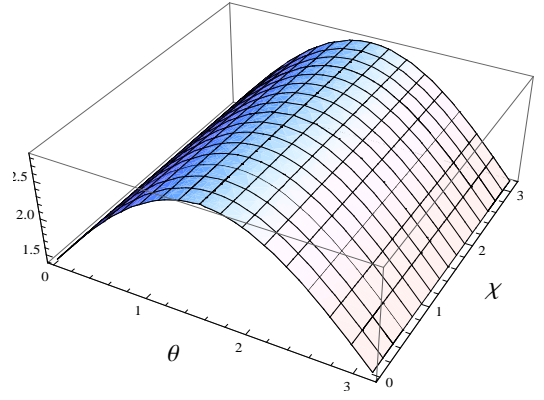} \caption{The numerically calculated
bound of $|\langle U^{\dag}W U\rangle_{\psi}|$ for the pure state
$\psi$. It is independent of the azimuthal angle $\chi$. }
\label{fig2}
\end{figure}
\begin{figure}[h]
\includegraphics[width=0.8\columnwidth,
height=0.6\columnwidth]{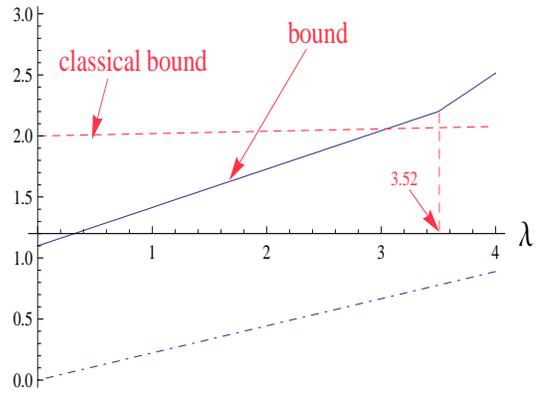} \caption{The bound of $|\langle
U^{\dag}W U\rangle_{\rho}|$ (solid line) as a function of $\lambda$
for the mixed state. When it is greater than the classical bound
(dashed line), we say that Alice and Bob can achieve the violation
of the CHSH inequality. The concurrence (dot-dashed line) of $\rho$
is also shown.} \label{fig3}
\end{figure}

In \cite{braunstein}, Braunstein \textit{et al.} showed that mixed
states can produce maximal violations of the CHSH inequality, and
the necessary and sufficient condition for violating the CHSH
inequality in an arbitrary mixed spin-$\frac{1}{2}$ state is
presented in \cite{horodecki}. Here we present a numerical method
which can be used to calculate the maximum of $|\langle
W\rangle_{\rho}|$ for any mixed state $\rho$ in the condition of
local measurement setting.

By using the eigenvectors of $W$ in Eq. (\ref{eigen}), we can
rewrite $|\langle U^{\dag} W U\rangle_{\rho}|_{max}$
($U=U^{a}\otimes U^{b}$) as
\begin{eqnarray}
|\langle U^{\dag} W
U\rangle_{\rho}|_{max}&=&\max_{U}\big|\bra{\eta_{1}}U\rho
U^{\dag}\ket{\eta_{1}}\bra{\eta_{1}}W\ket{\eta_{1}}\nonumber\\
&&+\bra{\eta_{3}}U\rho
U^{\dag}\ket{\eta_{3}}\bra{\eta_{2}}W\ket{\eta_{3}}\big|\nonumber\\
&=&
\max_{U}\Big|2\sqrt{2}\cdot\big[\tr(U^{\dag}\ket{\eta_{3}}\bra{\eta_{3}}U\rho)\nonumber\\
&&-\tr(U^{\dag}\ket{\eta_{1}}\bra{\eta_{1}}U\rho)\big] \Big| .
\label{x1}
\end{eqnarray}
Substituting $U^{a}\otimes U^{b}$, $\eta_{1}$, $\eta_{3}$ and $\rho$
into Eq. (\ref{x1}), we can numerically calculate the bound of
$|\langle U^{\dag} W U\rangle_{\rho}|$. In the case of pure state,
$\rho=\ket{\psi}\bra{\psi}$ where $\ket{\psi}$ is the state in Eq.
(\ref{state}), we calculate the bound by using the numerical scheme.
The results are shown in Fig. \ref{fig2}. They are the same as those
calculated from Eq. (\ref{outcome1}).

Now let us consider a mixed state which has a single positive
parameter
\begin{eqnarray}
\rho=\frac{1}{9}\left(\begin{array}{c} 1~~~~0~~~~0~~~~0\\
0~~~~4~~~~\lambda~~~~0\\
0~~~~\lambda~~~~4~~~~0\\
0~~~~0~~~~0~~~~0
\end{array}\right),
\label{q}
\end{eqnarray}
where we take $0\leq\lambda\leq4$ to ensure the positivity of
$\rho$. This state is usually used to investigate the evolution of
the entanglement of a pair of qubits exposed to local noisy
environments \cite{yu1,yu2}. We substitute Eq. (\ref{q}) into Eq.
(\ref{x1}) and calculate the bound of the Bell operator. The
obtained results are shown in Fig. \ref{fig3}. There is a turning
point of the curve near $\lambda=3.52$. From Fig. \ref{fig3} we can
see that the bound is highly consistent with the concurrence except
for this turning point. Due to the restriction of the local vertical
measurement scheme, only when the concurrence is greater than $0.6$
Alice and Bob can achieve the violation of the CHSH inequality. In
this form of mixed states the maximal violation of the CHSH
inequality cannot be realized. It is expected that only when the
concurrence of a mixed state is equal to $1$ the maximal violation
of the CHSH inequality could be achieved.

In summary, for any pure state we present an analytical expression
of the bound of the Bell operator in the condition that Alice and
Bob both perform local vertical measurements, and for a general
state we derive an numerical method for the calculation of the
bound. The results show intimate relationship between the bound and
the concurrence. We suggest that the bound of the Bell operator in
the condition of local vertical measurements may be used as a
measure of the entanglement.

\vskip 0.5 cm

{\it Acknowledgments} This work was supported by the State Key
Programs for Basic Research of China (Grant Nos.2005CB623605 and
2006CB921803), and by National Foundation of Natural Science in
China Grant Nos. 10474033 and 60676056.


\end{document}